\documentclass[preprint,aps,showpacs,amsmath,amssymb,amsfonts,floatfix,superscriptaddress]{revtex4}
\usepackage{graphicx}
\usepackage{dcolumn}
\usepackage{bm}
\usepackage{latexsym}
\usepackage{float}
\usepackage{supertabular}
\usepackage{longtable}

\begin{document}
\title{The Influence of Mobility Rate on Spiral Waves in Spatial Rock-Paper-Scissors Games}
\author{Mauro Mobilia}\email{M.Mobilia@leeds.ac.uk}
\affiliation{Department of Applied Mathematics, School of Mathematics, University of Leeds, Leeds LS2 9JT, U.K.} %
\author{Alastair M.~Rucklidge}
\affiliation{Department of Applied Mathematics, School of Mathematics, University of Leeds, Leeds LS2 9JT, U.K.}
\author{Bartosz Szczesny}
\affiliation{Department of Applied Mathematics, School of Mathematics, University of Leeds, Leeds LS2 9JT, U.K.}
\begin{abstract}
We  consider a two-dimensional model of three species in rock-paper-scissors  competition
and study the self-organisation of the population into fascinating spiraling patterns. Within our individual-based metapopulation formulation, the population composition changes due to cyclic dominance (dominance-removal and dominance-replacement), mutations, and pair-exchange of neighboring individuals. Here, we study the influence of mobility on the emerging patterns and investigate when the pair-exchange rate is responsible for spiral waves to become elusive in stochastic lattice simulations. In particular, we show that the spiral waves predicted by the system's deterministic partial equations are found in lattice simulations only within a finite range of the mobility rate. We also report that in the absence of mutations and dominance-replacement, the resulting spiraling patterns are subject to convective instability and  far-field breakup at low mobility rate. Possible applications of these resolution and far-field breakup phenomena are discussed.
\end{abstract}


\maketitle


\section{Introduction}
Understanding the mechanisms allowing the maintenance of species coexistence is an issue
of paramount importance~\cite{biodiversity}. 
Evolutionary game theory~\cite{games1,games2,games3}, where the success of one species depends on what the others are doing,
provides a fruitful framework to investigate this question by means of paradigmatic schematic models. In this context, cyclic dominance is considered  as a possible motif enhancing the maintenance of biodiversity, and models of populations in cyclic competition have recently received significant attention. 

The rock-paper-scissors (RPS) game - in which rock crushes scissors, scissors cut paper, and paper wraps rock - and its variants are paradigmatic models for the cyclic competition between three species. Examples of RPS-like dynamical systems  can be {\it Uta stansburiana} lizards, and communities of {\it E.coli}~\cite{PatternsRPS,Kerr,Nahum,LizardMutation}, as well as coral reef invertebrates~\cite{Jackson}. In the absence of spatial degrees of freedom and mutations, the presence of demographic fluctuations in finite populations leads to the loss of biodiversity with the extinction of two species in a finite time, see {\it e.g.}~\cite{RPSnonspatial1,RPSnonspatial2,RPSnonspatial3,RPSnonspatial4,RPSnonspatial5,RPSnonspatial7}. However, in nature, organisms typically interact with a finite number of individuals in their neighborhood and are able to migrate. It is by now 
well established both theoretically and experimentally that space and mobility greatly influence how species evolve and how ecosystems self-organize, see {\it e.g.}~\cite{Turing,Murray,Koch,patterns1,patterns2,patterns3}. Of particular relevance are the in vitro experiments with \textit{Escherichia coli} of Refs.~\cite{PatternsRPS,Kerr,Kerr2,Nahum} showing that, when arranged on a Petri dish, three strains of bacteria in cyclic competition coexist for a long time while two of the species go extinct when the interactions take place in well-shaken flasks. On the other hand,  in vivo experiments of Ref.~\cite{Kirkup} with bacterial colonies in the intestines of co-caged mice  have shown that mobility allows the bacteria to migrate between mice and to maintain their coexistence. These observations
illustrate that {\it mobility can both promote and jeopardize biodiversity in RPS games}, as argued in Refs.~\cite{RMF1,RMF2,RMF3}: In the experiments of Ref.~\cite{Kirkup} biodiversity is maintained only when  
there is migration, whereas in Ref.~\cite{Kerr} species coexistence is lost 
 in well-shaken flasks corresponding to a setting with a high mobility rate.

These considerations have motivated a series of studies aiming at investigating the relevance of spatial structure and individual's mobility on the properties of RPS-like systems. For instance, various two-dimensional versions of the model introduced by May and Leonard~\cite{MayLeonard}, characterized by cyclic ``dominance removal'' in which  each species ``removes'' another in turn (see below), have received much attention~\cite{RMF1,RMF2,RMF3,Matti,Jiang,He2,He3}. It has been shown that species coexist for a long time in these models with pair-exchange among neighboring individuals: below a certain mobility threshold species coexist by forming intriguing spiraling patterns below a certain mobility threshold, whereas there is loss of biodiversity 
above that threshold~\cite{RMF1,RMF2,RMF3}. 
 Another popular class of  RPS models are those characterized by a zero-sum cyclic interactions (via ``dominance replacement'') with
 a conservation law at mean-field level (``zero--sum'' games)~\cite{zerosum1,zerosum2,zerosum3,zerosum4,zerosum5,zerosum6,zerosum7,zerosum8,zerosum9} and whose dynamics in two-dimensions also leads to a long--lasting coexistence of the species but not to the formation of spiraling patterns, see {\it e.g.}~\cite{Matti,zerosum6,He3}. 
Recent studies, see {\it e.g.}~\cite{Rulands,RF08,SMR,SMR2,FigshareMovies,cycl-rev,BS}, have investigated the dynamics of two-dimensional RPS models combining cyclic dominance removal and replacement, while various 
generalization to the case of more than three species have also been considered, see {\it e.g.}~\cite{Avelino,Pleim1,Pleim2}. 
In  Refs.~\cite{SMR,SMR2,FigshareMovies,cycl-rev,BS} we studied the  spatio--temporal properties of a generic two--dimensional RPS-like model accounting for cyclic competition  with dominance--removal and dominance--replacement, along with other evolutionary processes such as reproduction, mutation and mobility via hopping and pair--exchange between nearest neighbors. By adopting a metapopulation formulation and using a multiscale and size-expansion analysis, combined with numerical simulations, we analyzed the properties of the emerging spatio--temporal dynamics. In particular, we derived the system's  phase diagram and characterized the spiraling patterns in each of the phases and showed how {\it non-linear mobility} can cause the far-field breakup of spiral waves. In spite of the predictions of the theoretical models, it is still unclear under which circumstances microbial communities in cyclic competition
would self-arrange into spiraling patterns as those observed in other systems such as myxobacteria and slime molds~\cite{spirals_exp,spirals_exp2}.

Here, we continue our investigation of the generic two--dimensional RPS-like model of Refs.~\cite{SMR,FigshareMovies,SMR2}
by focusing on the influence of pair-exchange between nearest-neighbors, as simplest form of migration, on the formation of spiraling patterns in two-dimensional lattice simulations. In particular, we demonstrate
a {\it resolution issue}: on a finite grid spiral waves can be observed only when the migration is within a certain range. We also show that in the absence of mutations and dominance-replacement, {\it e.g.} as in Refs.~\cite{RMF1,RMF2,RMF3,RF08}, the spiraling patterns emerging from the dynamics are subject to convective instability and  far-field breakup at low mobility rate.

This paper is structured as follows: The generic metapopulation model~\cite{metapopulation,FigshareMovies} is introduced in Sec.~2 and, building on Refs.~\cite{SMR,SMR2},  the main features of its description at mean-field level and in terms of partial differential equations (PDEs) are outlined. The section 3 is dedicated to a summary of the characterization of the system's spiraling patterns in terms of the underlying complex Ginzburg-Landau equation (CGLE).
Section 4 is dedicated to our novel results concerning the resolution issues in finite lattices 
and the far-field breakup and convective instability under low mobility. Finally, we conclude by summarizing and discussing our findings.

\section{Model}

\begin{figure}
\centering
	\includegraphics[width=0.425\linewidth]{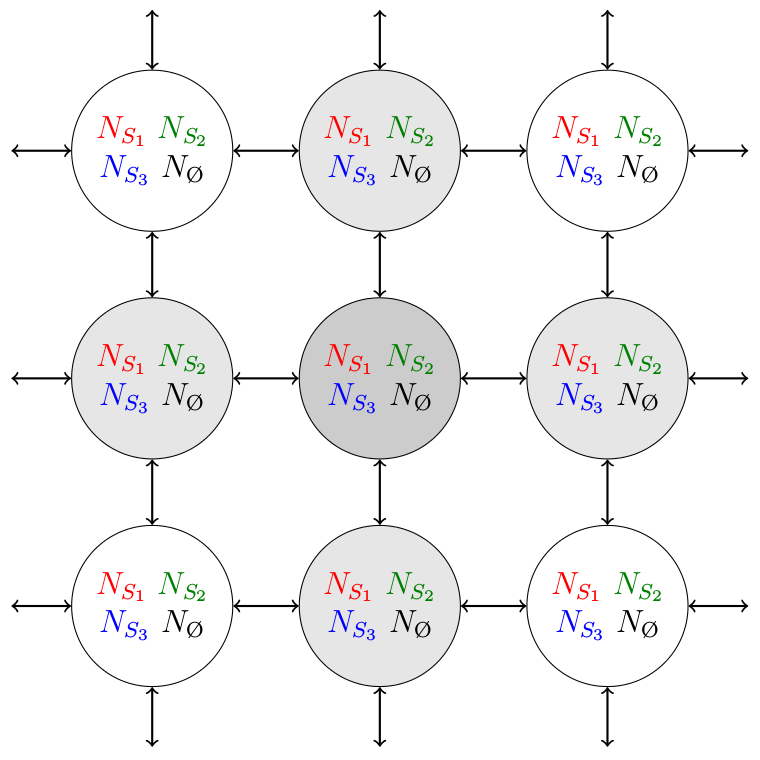}
	\caption{(Color online). Cartoon of the metapopulation model: $L \times L$ patches (or islands) are 
	arranged on a periodic square lattice (of linear size $L$). Each patch ${\bm \ell}=(\ell_1, \ell_2)$ can 
	accommodate at most $N$ individuals of species $S_1, S_2$, $S_3$ and empty spaces denoted $\emptyset$. 
	Each patch consists of a well--mixed population of $N_{S_1}$ individuals of species $S_1$, $N_{S_2}$ of type 
	$S_2$, $N_{S_3}$ of type $S_3$ and $N_{\emptyset} = N - N_{S_1} - N_{S_2} - N_{S_3}$ empty spaces. The 
	composition of a patch evolves in time according to the processes (\ref{sel}) and (\ref{non-sel}).
	Furthermore, migration from the focal patch (dark gray) to its four nearest--neighbors (light gray) 
	occurs according to the processes (\ref{migr}), see text. Adapted from \cite{SMR2}.}
	\label{Fig1}
\end{figure}

As in Refs.~\cite{SMR,SMR2}, we consider the generic model of cyclic dominance between three competing species defined on an $L\times L$ periodic 
square lattice of patches, $L$ being the linear size of the grid, where each node of the grid is labelled by a vector ${\bm \ell}=(\ell_1,\ell_2)$.
As illustrated in Fig.~\ref{Fig1}, each patch consists of a 
well-mixed population of species $S_1, S_2, S_3$ and empty spaces $\emptyset$ and has a limited carrying capacity $N$:
In each patch ${\bm \ell}$ there are therefore at most $N$ individuals
$N_{S_i}({\bm \ell})$  of species $S_i$ ($i=1,2,3$), and there are also $N_{\emptyset}({\bm \ell})=N-N_{S_1}({\bm \ell})-N_{S_2}({\bm \ell})- N_{S_3}({\bm \ell})$ empty spaces. Within each patch ${\bm \ell}$, the population composition evolves according to
the most generic form of cyclic RPS according 
to the following schematic reactions:
	\begin{eqnarray}
	\label{sel}
		S_i + S_{i+1} \xrightarrow{\sigma} S_i + \emptyset &~&
		S_i + S_{i+1} \xrightarrow{\zeta } 2 S_i \\
	\label{non-sel}
		S_i + \emptyset \xrightarrow{\beta} 2 S_i &~&
		S_i \xrightarrow{\mu} S_{i\pm1},
	\end{eqnarray}
where the  index $i \in \{1,2,3\}$ is ordered cyclically such that $S_{3+1} \equiv S_1$ and $S_{1-1} \equiv S_3$. 
The reactions (\ref{sel}) describe the generic form of cyclic competition,
that comprises {\it dominance--removal} with rate $\sigma$ and {\it dominance--replacement} with rate 
$\zeta$~\cite{SMR,SMR2,cycl-rev}. The processes (\ref{non-sel}) allow for the reproduction of 
each species (with rate $\beta$) independently of the cyclic interaction provided that free space ($\emptyset$) is 
available within the patch. The biological interpretation of the mutations $S_i \xrightarrow{} S_{i\pm1}$ (with rate $\mu$) is, {\it e.g.}, that they mimic the fact 
that side--blotched lizards {\it Uta stansburiana}  undergo throat--color transformations~\cite{LizardMutation}, while from a mathematical perspective they yield 
 a supercritical Hopf bifurcation at mean-field level, see Refs.~\cite{RPSnonspatial6,RPSnonspatial8}, about which a multiscale expansion is feasible, see below and Refs.~\cite{SMR,SMR2,cycl-rev}. Since we are interested in analyzing the spatio--temporal arrangement of the populations, in addition to the intra--patch reactions (\ref{sel})-(\ref{non-sel}), we also allow individuals to migrate between neighboring patches ${\bm \ell}$ and ${\bm \ell}'$ via pair exchange, according to 
	\begin{eqnarray}
	\label{migr}
		\big[X \big]_{{\bm \ell}} \big[Y \big]_{{\bm \ell}'} &\xrightarrow{\delta}&
		\big[Y	\big]_{{\bm \ell}} \big[X \big]_{{\bm \ell}'},
	\end{eqnarray}
where $X\neq Y\in \{S_1, S_2, S_3, \emptyset\}$. 

At an individual-based level, the model is defined by the Markov processes associated with the reactions (\ref{sel})-(\ref{migr}). The model's dynamics is thus described by the underlying master equation, and the stochastic lattice simulations performed using the Gillespie algorithm~\cite{Gillespie}, as explained in  Refs.~\cite{SMR,SMR2,BS}. 
The  metapopulation formulation of the model makes it well-suited for a size expansion of the master equation in $1/N$~\cite{SizeExp,Gardiner,TuringNoisy1,TuringNoisy2,SMR2,BS}.  Such an expansion in the 
 inverse of the  carrying capacity has been detailed in Ref.~\cite{SMR2} where we showed 
 that to lowest order in the continuum limit ($L\gg 1$) on a square domain of size ${\cal S}\times {\cal S}$
 the system evolves according to the partial differential equations  (with periodic boundary conditions) 
	\begin{eqnarray}
	\label{pde}
		\partial_t s_i
		&=& D \Delta s_i +s_i[1-\rho - \sigma s_{i-1}] + \zeta s_i[s_{i+1} - s_{i-1}]+
		\mu \left[s_{i-1} + s_{i+1} - 2 s_i\right]
	\end{eqnarray}
 for the species densities $s_i=N_{S_i}/N\equiv s_i({\bm x},t)$, where ${\bm x}={\cal S} {\bm \ell}/L$ is a continuous variable, and $\rho=s_1+s_2+s_3$. Here and henceforth, without loss of generality we have rescaled time by setting $\beta =1$, and on average there are $N$
microscopic interactions during a unit of time~\cite{SMR2}.
As usual, the diffusion coefficient $D$ and the migration rate $\delta$ are simply related by $D=\delta ({\cal S}/L)^2$.\\
It is worth noting that contrary to 
the models of Refs.~\cite{SMR,SMR2}, here we do not consider non-linear diffusion: Movement in  (\ref{pde}) appears 
only through the {\it linear diffusive terms $D \Delta s_i $}. Furthermore, while in Refs.~\cite{SMR,SMR2} we 
focused on  spatio-temporal patterns whose size exceeds that of  the lattice unit spacing
and we mostly considered domains 
of size ${\cal S}=L$ so that $D=\delta$, here we prefer to keep ${\cal S}$ and $L$ separate, and therefore  $D$ and $\delta$, distinct.
Eqs.~(\ref{pde}) are characterized by an interior fixed point ${\bm s}^*=(s_1^*,s_2^*,s_3^*)$ 
associated with the coexistence of the three species with the same density $s_i^*=s^*=1/(3 + \sigma)$.
In the absence of space (i.e. upon setting $\Delta s_i=0$ in (\ref{pde})) and with no mutations ($\mu=0$),
 ${\bm s}^*$ is never asymptotically stable and the 
 mean field dynamics yields heteroclinic cycles (when $\mu=0, \sigma>0$ and $\zeta\geq 0$, 
 with $\zeta=0$ corresponding to the degenerate case)~\cite{MayLeonard}
 or neutrally stable periodic orbits (when $\mu=\sigma=0$ and $\zeta>0$)~\cite{games1,games2,games3} and finite-size fluctuations 
always lead to  the quick extinction of two species~\cite{RPSnonspatial1,RPSnonspatial2,RPSnonspatial3,RPSnonspatial4,RPSnonspatial5}. However, quite interestingly when 
the mutation rate is non-zero, at mean-field level a supercritical Hopf bifurcation (HB) occurs
at $\mu_H=\sigma/[6(3 + \sigma)]$ and yields a stable limit cycle when $\mu<\mu_H$~\cite{SMR} (see also Refs.~\cite{RPSnonspatial6,RPSnonspatial8}).

\begin{figure}
	\centering
	\includegraphics[width=0.8\linewidth]{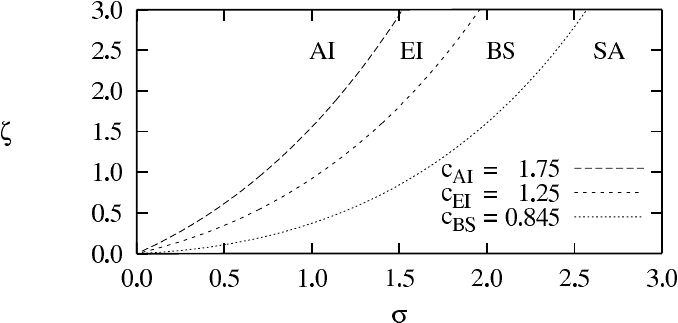}
	\caption{ Phase diagram of the two--dimensional RPS system around the Hopf bifurcation with contours of $c = (c_{{\rm AI}}, c_{{\rm EI}}, c_{{\rm BS}})$ in the $\sigma-\zeta$ plane, see text. We distinguish four phases: spiral waves are unstable in AI, EI and SA phases, while they are stable in BS phase. The boundaries between the phases have been obtained using the CGLE parameter (\ref{c3}). Adapted from \cite{SMR2}.}
 	\label{diagram_phases}
\end{figure}
\section{Spiraling patterns and the complex Ginzburg-Landau equation}
In Refs.~\cite{SMR,SMR2} we showed that  the dynamics in terms of the PDEs (\ref{pde}) yield spiraling patterns
whose spatio-temporal properties can be analyzed in terms of the system underlying complex Ginzburg-Landau equation (CGLE)~\cite{CGLE}. 

The latter is derived  by introducing the
``slow variables'' $({\bm X}, T)=(\epsilon{\bm x},~\epsilon^2 t)$, where $\epsilon=\sqrt{3(\mu_H-\mu)}$ is the system's small parameter in terms of which a multiscale 
expansion is performed about the HB~\cite{Miller}. Details of the derivation can be found in Ref.~\cite{SMR2}
and brief accounts in Refs.~\cite{SMR,cycl-rev}. Here we quote the
system's CGLE for the complex modulated amplitude $\mathcal{A}({\bm X}, T)$ which is a linear combination of the rescaled species 
densities~\cite{SMR,SMR2,cycl-rev}:
	\begin{equation}
	\label{CGLE}
		\partial_{T} \mathcal{A} =
		D \Delta_{\bm X} \mathcal{A} + \mathcal{A} - (1 + i c) |\mathcal{A}|^2 \mathcal{A},
	\end{equation}
where  $\Delta_{\bm X}= \partial_{X_1}^2+\partial_{X_2}^2=\epsilon^{-2}(\partial_{x_1}^2+\partial_{x_2}^2)$ and 
$\partial_{T}=\epsilon^{-2}\partial_{t}$, and after having rescaled $\mathcal{A}$ by a constant, we find the parameter
	\begin{equation}
	\label{c3}
		c = \frac
		{12\zeta (6 - \sigma)(\sigma + \zeta) + \sigma^2 (24 - \sigma)}
		{3\sqrt{3} \sigma (6 + \sigma)(\sigma + 2\zeta)}.
	\end{equation}

\begin{figure}
	\centering
	\includegraphics[width=0.764\linewidth]{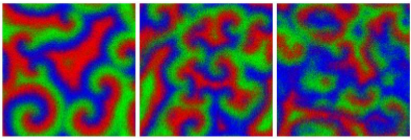}
	\caption{
(Color online).  Typical long-time snapshots in the BS (left), EI (middle) and AI (right) phases
from stochastic simulations  at low mutation rate. As in the next figures, each color represents one species (dark dots are regions of low density).
 The parameters are $L=128$, $N=64$, $(\beta,\sigma,\mu,\delta)=(1,1,0.001,1)$, and $\zeta=0.6$ (left),
$1.2$ (middle), and $\zeta=1.8$ (right), see text. In all panels, the initial condition is a random perturbation of the homogeneous state 
$\bm{s}^*$, see text and \cite{FigshareMovies}. Adapted from \cite{BS}.
 }
 	\label{snapshots}
\end{figure}

As explained in Refs.~\cite{SMR,SMR2}, the CGLE (\ref{CGLE}) allows us to accurately characterize the spatio-temporal spiraling patterns in the vicinity of the HB (for $\epsilon\ll 1$ i.e. $\mu\lesssim \mu_H$) by using the well-known phase diagram of the two-dimensional CGLE, see {\it e.g.}~\cite{CGLE}, and
to gain significant insight into the system's spatio--temporal behavior away from the HB (we here restrict $\sigma$ and $\zeta$ into [0,3]):
\begin{itemize}
\item For $\mu\lesssim \mu_H$ (close to the HB)~\cite{SMR}: There are four phases separated by the three critical values $(c_{{\rm AI}}, c_{{\rm EI}}, c_{{\rm BS}})\approx (1.75, 1.25, 0.845)$, as  shown in the phase diagram of Fig.~\ref{diagram_phases}: No spiral waves can be sustained in the 
 ``absolute instability (AI) phase'' ($c>c_{{\rm AI}}\approx 1.75$); spiral waves are  convectively unstable
in the Eckhaus instability (EI) phase with $c_{{\rm EI}}\approx 1.25<c<c_{{\rm AI}}$; stable spiral waves
 are found in 
the bound state (BS) phase ($c_{{\rm BS}}\approx 0.845<c<c_{{\rm EI}}$); while spiral waves 
annihilate
when they collide in the spiral annihilation (SA) phase 
 when $0<c<c_{{\rm BS}}$.
\item For $\mu\ll \mu_H$ (away from the HB)~\cite{SMR2}:
 The AI, EI and BS phases are still present even away from the HB whose boundaries
are essentially the same as in the vicinity of the HB, see Fig.~\ref{snapshots}. At low mutation rate, there
is no spiral annihilation and the SA phase is generally replaced by an extended BS phase
(with  far-field breakup of the spiral waves when $\sigma \gg \zeta$).
\end{itemize}

Away from the cores of the spiral waves, the solution to the CGLE (\ref{CGLE})
can be approximated by the travelling-wave ansatz $\mathcal{A}({\bm X},T)=Re^{i({\bm k}\cdot{\bm X}-\omega T)}$ 
of amplitude $R(c)$, angular frequency $\omega$ and wave number ${\bm k}$~\cite{SMR2}. As detailed in Ref.~\cite{SMR2}, 
the wavelength $\lambda=2\pi/(\epsilon k)$ of the spiraling patterns in the BS and EI phases near the HB in the physical space (in lattice units) can thus be estimated
by $\lambda\approx \lambda_H$, where 
\begin{eqnarray}
\label{lambda}
\lambda_H=\frac{2\pi L}{\epsilon {\cal S}}
\sqrt{\frac{D}{1-R^2(c)}}=\frac{2\pi}{\epsilon}
\sqrt{\frac{\delta}{1-R^2(c)}},
\end{eqnarray}
where the amplitude $R^2(c)$ is a decreasing function of $c$ in the BS and EI phases (see Fig.~6 in \cite{SMR2}) and has been found numerically to be in the range 0.84-0.95
in the BS phase~\cite{SMR2}.

In Ref.~\cite{SMR2}, we showed that the description in terms of the CGLE 
is not only valid and accurate near the HB ($\mu\lesssim \mu_H$), but it is also insightful
in the limit $\mu\ll \mu_H$ of low mutation rate. 
In fact, we showed that lowering the mutation rate $\mu$
below $\mu_H$ results in three regimes: the AI, EI and BS phases, see Fig.~\ref{snapshots}. We also found that  reducing $\mu$ below $\mu_H$ results into
shortening the wavelength $\lambda$ of the spiraling patterns in the BS and EI phases:
 The wavelength at low mutation rate satisfies a linear relationship (see Fig.~14 in Ref.~\cite{SMR2}): $\lambda \approx \lambda_{\mu\ll \mu_H}$, where
\begin{eqnarray}
\label{lambd}
\lambda_{\mu\ll \mu_H}=(\lambda_{H} - \lambda_{0})\frac{\mu}{\mu_H} + \lambda_0,
\end{eqnarray}
where the wavelength $\lambda_{0}$ at $\mu=0$ is inferred from the numerical solution of the PDEs (\ref{pde})
and is shorter than the wavelength $\lambda_{\mu_H}$ near ${\mu_H}$;
 typically
$\lambda_{0}\in [0.3\lambda_{\mu_H},0.5\lambda_{\mu_H}]$ and it scales as $\lambda_{0}\sim \sqrt{2\delta(3+\sigma)/\sigma}$~\cite{Alastair}.
For instance, when $(\beta,\sigma,\zeta,\mu,\delta)=(1,1,0.6,0.01,0.64)$, we have $\epsilon\approx 0.308$, $R^2\approx 0.9$
and (\ref{lambda}) yields $\lambda_H \approx 52$ while we found $\lambda_{0}\approx 26$ 
and therefore at  $\mu=0.01$,  the wavelength is $\lambda_{\mu=0.01} \approx 32$ which is in good agreement with lattice simulations, see Fig.~\ref{resolution} (bottom, left).

\section{How does pair-exchange influence the formation of spiral waves  on a grid?}
We have seen that near the HB the RPS dynamics is generally 
well described in terms of the PDEs (\ref{pde}) and CGLE (\ref{CGLE})
when $N\gg 1$. Accordingly, the effect of space and individuals' mobility 
is accounted by linear diffusion. In this setting, 
rescaling the mobility rate $\delta \to \alpha \delta$ ($\alpha >0$)
boils down to rescale the diffusion coefficient and space according to
$D\to \alpha D$ and ${\bm x} \to {\bm x}/\sqrt{\alpha}$. Hence, the size of the 
spatial patterns increases  when the individuals' mobility is increased,
and it decreases when the mobility is reduced~\cite{RMF1,RMF2,RMF3,SMR2}.
Furthermore, according to the description in terms of the CGLE, the mobility rate and diffusion coefficient do 
not affect the stability of the spiraling patterns near the HB but only change their size.
 Below, we discuss the cases of very small or large
mobility rate and show that this may result in 
 spiral waves being elusive and/or unstable even in  regimes where the PDEs (\ref{pde}) and  CGLE (\ref{CGLE})
predict that they would exist and be stable.
\subsection{Resolution issues}
\begin{figure}[H]
	\centering
	\includegraphics[width=0.35\linewidth]{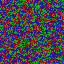}
        \includegraphics[width=0.35\linewidth]{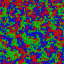}
\includegraphics[width=0.35\linewidth]{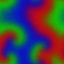}
\includegraphics[width=0.35\linewidth]{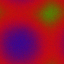}
	\caption{Typical snapshots of stochastic simulations of the model with $(\beta,\sigma,\zeta,\mu)=(1,1,0.6,0.01)$, $N=1024$ and $L=64$ and
 different mobility rates after a time 
$t=200$. 
The  mobility rate and predicted wavelength are from left to right: $(\delta, \lambda)=(0.000625,1)$ and  
$(\delta, \lambda)=(0.005625,3)$ in the top row, and $(\delta,\lambda)=(0.64,L/2)$ and  $(\delta,\lambda)=(1.44, 3L/4)$ in the bottom row, see text. 
Adapted from \cite{BS}.}
 	\label{resolution}
\end{figure}

In this section, we focus on resolution issues and show that while the description
of the dynamics in terms of the PDEs (\ref{pde}) predict the formation of spiral waves these cannot be observed on finite lattice due to resolution issues. In other words, we
report that, even when the PDEs (\ref{pde}) and  CGLE (\ref{CGLE}) predict that the dynamics leads to stable spiraling patterns (BS phase), these 
 may be elusive when the mobility rate is too low or too high as illustrated in Figure~\ref{resolution}.

In order to determine the
 range of the mobility rate $\delta$ within which stable spiral waves can be observed 
on a two-dimensional grid, we distinguish three regimes, see Table~\ref{Table1}: (i) $\lambda\sim o(1)$;
(ii) $1\ll \lambda \ll L$; (iii)~$\lambda \gtrsim {\cal O}(L)$.
In regime (i), the PDEs (\ref{pde}) 
predict a myriad of tiny spirals of wavelength of the order of one unit lattice space. Clearly, the resolution of any finite lattice is insufficient to allow to observe spiraling patterns of such a tiny size (of order of one pixel) on the grid, see Figure~\ref{resolution} (top). In this situation, instead of spiral waves lattice simulations lead to apparently clumps of activity. As shown in  Figure~\ref{resolution}, this phenomenon does not stem from demographic noise since it is  present even when $N$ is very large, as shown in  Figure~\ref{resolution} (where $N=1024$). Yet, due to their small size, these emerging incoherent spatio-temporal structures are prone to be affected by demographic fluctuations and result in being noisy. In regime (ii), the 
 spiral waves' wavelengths, $\lambda_H$ (\ref{lambda}) near the HB and $\lambda_{\mu\ll \mu_H}$ (\ref{lambd})
 at low mutation rate, are much larger than the inter-patch space and smaller than the domain size. Hence, stable spiraling patterns fit within the lattice and are similar to those predicted 
by the PDEs~(\ref{pde}), see Figure~\ref{resolution}~(bottom left). In regime (iii), the predicted $\lambda_H$ and $\lambda_{\mu\ll \mu_H}$ outgrow the lattice and the arms of the resulting large spirals  appear like planar waves, see Figure~\ref{resolution}~(bottom right). Interestingly, planar waves have been found in the model of Ref.~\cite{Rulands} without mutations ($\mu=0$)  at sufficiently high pair-exchange rate.

In order to estimate the boundaries between these regimes, we can 
use the \mbox{relations 
(\ref{lambda}) and (\ref{lambd}}) in the BS phase. According to those, 
$\lambda= \kappa \sqrt{\delta}/\epsilon$, where
 $\kappa=2\pi/\sqrt{1-R^2(c)}$ is a constant such $15.7\lesssim \kappa\lesssim 28.1$.
Hence, the regime (i) corresponds to low mobility rates of order $\delta \sim o(\epsilon^2)$;
in regime (ii) we have $ \epsilon^ 2\ll \delta \lesssim (L\epsilon/\kappa)^2$ (intermediate mobility rate); while in regime (iii) $\delta \gtrsim (L\epsilon/\kappa)^2$ (high mobility rate),  as summarized in the following table where we have also 
included the corresponding diffusion coefficient $D=\delta/L^2$ on a domain of unit size (${\cal S}=1$):

\begin{table}[H]
\caption{Spatio-temporal patterns emerging in three different regimes, at low (i), intermediate (ii) and high (iii) mobility rate  (top to bottom).}
\vspace{3mm}
\centering
 \begin{tabular}{ c|c|c|c }
\hline
  {\it \textbf{Wavelength $\lambda$}} & {\it \textbf{Mobility Rate $\delta$}}  & {\it \textbf{Diffusion Coefficient $D$ (${\cal S}=1$)}} & {\it \textbf{Patterns on Grid}} \\ \hline 
  $\lambda \sim  {\cal O}(1)$ & $\delta \sim o(\epsilon^2)$ & $D \sim o((\epsilon/L)^2)$ & Clumps of activity \\ 
    $1 \ll \lambda \lesssim L$ &  $ \epsilon^ 2\ll \delta \lesssim (L\epsilon/\kappa)^2$  & $ 
(\epsilon/L)^ 2\ll D \lesssim (\epsilon/\kappa)^2$  & Stable spirals \\ 
  $\lambda \gtrsim L$ & $\delta \gtrsim (L\epsilon/\kappa)^2$ &  $D \gtrsim (\epsilon/\kappa)^2$ & Planar waves  
\label{Table1}
\\
\hline
\end{tabular}
\end{table}

This means that stable spirals of wavelength given by (\ref{lambda}) or (\ref{lambd})
can be observed in the BS phase in the range of mobility rate $\epsilon^ 2\ll \delta \lesssim (L\epsilon/\kappa)^2$ that grows with $L$. Hence, when $L$ is sufficiently large lattice simulations will lead to the formation of stable spiral waves almost for any finite 
mobility rate. However, as the size of plates used in most microbial experiments rarely exceeds $L=100$~\cite{Kerr}, it is interesting to consider the case where $L$ is not too large. In particular, when the ratio $L/\kappa={\cal O}(1)$, the range $\epsilon^ 2\ll \delta \lesssim (L\epsilon/\kappa)^2$ is finite and spiral waves outgrow the lattice even 
for a finite mobility rate $\delta \gtrsim (L\epsilon/\kappa)^2$.
 For instance, in Figure~\ref{resolution}, we have $L=64$ and
$(\beta,\sigma,\zeta,\mu,\delta, N, L)=(1,1,0.6,0.01, 1024)$ which yields $\epsilon\approx 0.3$, $c\approx 1.0$, $R^{2}\approx 0.9$~\cite{SMR2,BS}, and $L\epsilon/\kappa \approx 0.92$. In this example,  with (\ref{lambda}) and (\ref{lambd}) we find
$\lambda(\delta =0.64)=L/2$ and  $\lambda(\delta=1.44)=3L/4$. According to the above discussion
we expect to find visible spiral waves for $\delta =0.64$ and patterns resembling planar waves
when $\delta=1.44$ which is confirmed by the bottom row of
Figure~\ref{resolution}. For the example of Figure~\ref{resolution}, the PDEs (\ref{pde}) predict
the formation of small spiral waves of wavelength $\lambda=1$ and $\lambda=3$  for $\delta=0.000625$ and 
 $\delta=0.005625$ respectively, which result in the noisy clumps of activities on the lattice of the top panel of Figure~\ref{resolution}.
\subsection{Far-field breakup of spiral waves under weak pair-exchange rate}
\begin{figure}
	\centering
	\includegraphics[width=0.8\linewidth]{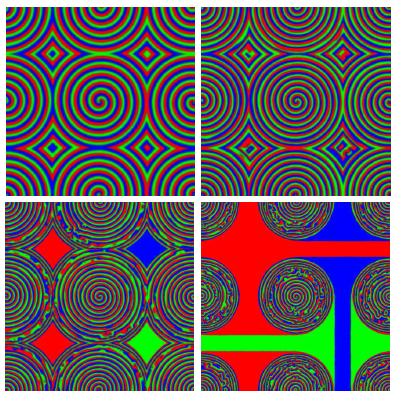}
	\caption{
(Color online). Typical snapshots of stochastic lattice simulations of the model with $(\beta,\sigma,\zeta,\mu)=(1,1,0,0)$, $N=256$ and $L=512$ and different mobility after a time 
$t=800$. In each panel, the initial condition (geometrically ordered and partially visible in the bottom right panel) is the same. 
The  mobility rate is from left to right: $\delta=0.4$ and  $\delta=0.2$ in the top row, and $\delta=0.1$ and  $\delta=0.05$ in the bottom row. Adapted from \cite{BS}.}
 	\label{farfield}
\end{figure}
Variants of the two-dimensional RPS model (\ref{sel})-(\ref{migr}) without mutation ($\mu=0$) have received significant interest and  many authors have studied under which circumstances the dynamics leads to the formation of stable spiraling patterns, see, {\it e.g.}, \cite{zerosum1,zerosum4,zerosum5,zerosum6,zerosum7,RMF1,RMF2,RMF3,RF08,Matti,He2}. In Ref.~\cite{RMF1,RMF2,RMF3}, where only the dominance-removal was considered, it was found that the cyclically competing populations moving under pair-exchange always form persisting spiraling patterns under a critical mobility threshold whereas no such coherent patterns were found  in a similar system where the cyclic competition was implemented according to the dominance-replacement process, see {\it e.g.}~\cite{Matti,zerosum6,He3}. By means of  an approximate mapping onto a CGLE,  the authors of Ref.~\cite{Matti} concluded that while the model with dominance-removal ($\sigma>0,~\zeta=0$) can sustain spiral waves, this is not the case of models with dominance-replacement ($\sigma=0,~\zeta>0$). In Ref.~\cite{RF08}, it is found that the combination dominance-removal and dominance-replacement processes can lead to stable spiraling patterns as well as to convectively unstable spiral waves. This picture was complemented and unified in our recent work~\cite{SMR,SMR2,cycl-rev,BS} where these questions were considered in the presence/absence of a small mutation rate and {\it nonlinear mobility} (pair-exchange and hopping processes were divorced). In particular, we showed that when the mutation rate is low or vanishes, {\it nonlinear mobility} alters the stability of the spiral waves and is responsible for their far-field breakup.

In this section we report that a similar intriguing phenomenon also occurs in the case where 
the mobility of the individuals is implemented by the simple nearest-neighbor
 pair-exchange~(\ref{migr}) which result in  {\it linear diffusive terms} in the corresponding PDEs~(\ref{pde}). To  characterize this novel phenomenon we have implemented the cyclic dominance  by considering only 
dominance-removal, i.e. we have set $\sigma=\beta=1$ and $\zeta=\mu=0$, and let the pair-exchange rate $\delta$ vary. This variant of the model is therefore the metapopulation version (here $N=1024$) of the model considered {\it e.g.} in \cite{RMF1,RMF2,RMF3,Matti} (where $N=1$).
Based on these previous works, see, {\it e.g.} \cite{RMF1,RMF2,RMF3,Matti,SMR,SMR2}, we would anticipate that the dynamics of such a variant of the model would be characterized by the formation of stable spiral waves. As shown in Fig.~\ref{farfield} (top, left), this is indeed the case when the pair-exchange is sufficiently high ($\delta=0.4$). However, when $\delta$ is lowered the
spiral waves become {\it far--field unstable}, see Fig.~\ref{farfield}, after the shortening of their  wavelength according to the scaling $\lambda\sim \sqrt{\delta}$. Hence, as the wavelength is reduced under low mobility, it appears that the core of the spirals can sustain its arms only for a few rounds before a convective instability starts growing, very much like in the EI phase, and eventually cause the far-field breakup of the spiral waves. While the detailed mechanism  of this far-field breakup has still to be elucidated, we believe that
 it stems from the nonlinear nature of the problem since  demographic noise is not at its origin (for $N=1024$ fluctuations are here negligible). We also think that a spiral far-field breakup always arises when $\zeta\approx 0$ and $\mu=0$, but depending on the mobility rate $\delta$ it occurs outside the lattice (high mobility rate) or within the grid  (low mobility rate). In fact, a careful analysis of  the PDEs (\ref{pde}) explains this phenomenon of far-field breakup with $\mu=0$: For fixed $\sigma>0$, spiral waves exhibit far-field breakup when $\zeta \gtrsim \sigma/2$ or when $\zeta$ is close to zero.
 However, this analysis is beyond the scope of this paper and will be given elsewhere~\cite{Alastair}.

. 

\section{Summary \& Conclusion}

We have studied the influence of a simple form of mobility, modeled by a pair-exchange between nearest neighbors,
on the spatio--temporal patterns emerging in the generic two-dimensional model of Refs.~\cite{SMR,SMR2} 
for the  cyclic rock-paper-scissors competition between three species. The underlying evolutionary processes are  cyclic dominance--removal and cyclic dominance--replacement interactions, reproduction, migration (pair-exchange), and mutation. While various properties of this system, formulated as a metapopulation model, were investigated in Refs.~\cite{SMR,SMR2} by combining multiscale and size expansions with numerical simulations, here we have analyzed the influence of the simple pair-exchange process on the properties of the spiraling patterns characterizing the dynamics of this system.

First, we have highlighted resolution issues that arise on finite lattices: While the description of the dynamics in terms of the underlying partial differential equations predict the formation of spiraling patterns in the so-called ``bound-state phase'', spiral waves may simply be elusive in the simulations on a finite lattice if the mobility rate is too low. More precisely, when the size of the lattice is finite and comparable to the size of experimental plates, we show that well-defined spiral waves can be observed only when the mobility rate is within a finite range: When the mobility rate is too low, the PDEs predict the emergence of spiral waves of wavelength of order of the lattice spacing which cannot be resolved, whereas when the mobility rate is sufficiently high the 
resulting spiraling patterns have a wavelength of the order of the lattice size and appear like planar waves.  
Spiral waves can be observed in lattice simulations when the mobility rate is between these two values.
In fact, building on the analysis carried out in Refs.~\cite{SMR,SMR2} in terms of the system's complex Ginzburg-Landau equation, we have estimated the critical value of the mobility rate. While the range within which spiral waves in bound-state phase grows with the size of the system, we have found that such a range may be finite and therefore spiraling patterns  too small to be resolved and observable on lattices of a size comparable to the plates used in most experiments (96-well plates, see {\it e.g.} ~\cite{Nahum}). We believe that these ``resolution issues'' may therefore be particularly relevant when one tries to interpret experimental results and can explain why spiral waves appear to be elusive in microbial experiments as those of \cite{Kerr,Nahum}.

Second, we have focused on  the version of the model with cyclic dominance--removal and without mutations that has received significant attention in the recent years, see {\it e.g.}~\cite{RMF1,RMF2,RMF3}. While  previous works reported that in this case the underlying rock-paper-scissors dynamics  (with $\zeta=\mu=0$) leads to well-defined spiraling patterns, here we show that spiraling patterns become convectively unstable and that a far-field breakup occurs when the mobility rate is lowered (with the other rates kept constant).  We have verified that the mechanism underlying this phenomenon  does not originate from demographic fluctuations and refer to the future work for a detailed analysis of its mechanism in terms of the system's partial differential equations~\cite{Alastair}.

While we have specifically discussed the biologically-relevant case of the two-dimensional  metapopulation model, this analysis can be readily extended to one-dimensional and three-dimensional lattices on which we would respectively expect traveling and scroll waves instead of spiral waves. It would also be interesting to study whether similar phenomena would arise to the oscillating patterns characterizing some RPS games on small-world networks~\cite{OnNetworks1,OnNetworks2,OnNetworks3}.

\vspace{6pt} 


\section{Acknoweldgments}
The support to BS via an EPSRC PhD studentship (Grant No. EP/P505593/1)  is gratefully acknowledged.

\vspace{6pt} 

The authors declare no conflict of interest.

\end{document}